\documentclass[aip,rsi,twoside,twocolumn,showpacs,reprint,amsmath,amssymb,preprintnumbers,superscriptaddress,floatfix]{revtex4-1}
\usepackage{textpos}
\usepackage{dcolumn}
\usepackage{bm}
\usepackage{cases}
\usepackage{upgreek}
\usepackage{txfonts} 

\usepackage{ifpdf}
\ifpdf
   \usepackage[pdftex]{graphicx}
   \usepackage[pdftex, pdfpagelayout=TwoColumnRight, pdfstartview=FitH, plainpages=false, pdfpagelabels]{hyperref}
  \else
    \usepackage[dvips]{graphicx}
\fi 

\begin{document}


\title{Membrane-based nanocalorimeter for high-resolution measurements of low-temperature specific heat}

\author{S. Tagliati}
\author{V. M. Krasnov}
\author{A. Rydh}
\email{andreas.rydh@fysik.su.se}
\affiliation{Department of Physics, Stockholm University, AlbaNova University Center, SE -- 106 91 Stockholm, Sweden}%

\date{\today}

\begin{abstract}
A differential, membrane-based nanocalorimeter for general specific heat studies of very small samples, ranging from $0.5\,\mathrm{mg}$ to sub-$\upmu\mathrm{g}$ in mass, is described. The calorimeter operates over the temperature range from above room temperature down to $0.5\,\mathrm{K}$. It consists of a pair of cells, each of which is a stack of heaters and thermometer in the center of a silicon nitride membrane, in total giving a background heat capacity less than $100\,\mathrm{nJ/K}$ at $300\,\mathrm{K}$, decreasing to $10\,\mathrm{pJ/K}$ at $1\,\mathrm{K}$. The device has several distinctive features: i) The resistive thermometer, made of a $\mathrm{Ge}_{1-x}\mathrm{Au}_{x}$ alloy, displays a high dimensionless sensitivity $\left| \mathrm{dln}R/\mathrm{dln}T \right| \gtrsim 1$ over the entire temperature range. ii) The sample is placed in direct contact with the thermometer, which is allowed to self-heat. The thermometer can thus be operated at high dc current to increase the resolution. iii) Data are acquired with a set of eight synchronized lock-in amplifiers measuring dc, $1^{\mathrm{st}}$ and $2^{\mathrm{nd}}$ harmonic signals of heaters and thermometer. This gives high resolution and allows continuous output adjustments without additional noise. iv) Absolute accuracy is achieved via a variable-frequency-fixed-phase technique in which the measurement frequency is automatically adjusted during the measurements to account for the temperature variation of the sample heat capacity and the device thermal conductance. The performance of the calorimeter is illustrated by studying the heat capacity of a small Au sample and the specific heat of a $2.6\,\upmu\mathrm{g}$ piece of superconducting Pb in various magnetic fields.
\end{abstract}


\maketitle

\section{INTRODUCTION}
Accurate thermodynamic measurements are essential to understand fundamental properties of materials in various fields of physics. In condensed matter, the measurement of  specific heat  is a central characterization applicable to all kind of materials. Low-temperature calorimetry is particularly suited for the investigation of superconductors and other novel systems with electronic phase transitions.\cite{Stewart:1983tp,Carbotte:1990zz,Schilling:1996ce} Such measurements require high resolution, since the electronic contribution to the heat capacity is only a minor part of the total heat capacity, except at the very lowest temperatures. The high resolution can be achieved through differential calorimeter designs and various temperature-modulated techniques.\cite{Sullivan:1968uo,Bachmann:1972vh,Graebner:1989jw,Hatta:1997vl,Kraftmakher:2002cv} Good accuracy is also often needed. The temperature dependence of the specific heat may, for instance, reveal central aspects of the nature of the electronic system, including energy gap structure, anisotropy, and possible signatures of quantum phase transitions. Measurements are, furthermore, often performed in magnetic fields of various direction, using single crystals of highest available quality. This requires both a small calorimeter and small samples. All this together puts high demands on the calorimeter setup, typically excluding  commercially available calorimeters. To meet this demand, calorimetry development is going in the direction of nanocalorimetry. Nanocalorimetry is a rapidly growing area of research, driven by several fields of physics. Nanocalorimeters include absorption sensors,\cite{Caspary:1999tq} devices to study transition enthalpies,\cite{Nakagawa:1998uj} fast scanning calorimeters to study microscopic nanostructure ensembles and thin films\cite{Efremov:2000vo,Efremov:2004bf,Lopeandia:2005vn} and kinetics and glass transitions of polymers\cite{Minakov:2005vh,Minakov:2007ix}, and combinatorial calorimeters.\cite{McCluskey:2010be} There are also several low-temperature microcalorimeters for $\mathrm{mg}$ samples\cite{Bachmann:1972vh,Schilling:1995wc,Schnelle:1995ti,Wilhelm:2004hw,Tokiwa:2011th} and nanocalorimeters for heat capacity measurements of thin films\cite{Denlinger:1994wf,Queen:2009bb} and very small ($\upmu\mathrm{g}$) samples at high\cite{Guenther:2011uz}, intermediate,\cite{Riou:1997tq,Lortz:2005hr,Minakov:2005jd,Rydh:2006ta,Marone:1997uz,Garden:2009eu,Lopeandia:2010kj} and down to low temperatures.\cite{Graebner:1989jw,Bourgeois:2005jw,Fon:2005dk,Cooke:2008fg} 

Going down in sample size favors the use of temperature-modulated techniques with corresponding high resolution, but makes it harder to obtain good absolute accuracy. This is partly due to an increasing relative contribution of the device addenda, but also due to non-adiabatic conditions and practical design issues, such as thermometry, system complexity, and thermal links considerations. Devices for general use that combine high-resolution measurements of small samples with good absolute accuracy over an extended temperature range are hard to find. The membrane-based nanocalorimeter presented here is intended to fill this gap. 

Our nanocalorimeter is developed for measurements of the temperature- and field dependence of the absolute specific heat of samples with a typical mass around $0.1$--$20\,\upmu$g, and for angular-dependent studies in magnetic fields. The device is built onto a pair of silicon nitride membranes using thin film techniques that provide low background heat capacity, less than $100\,\mathrm{nJ/K}$ at $300\,\mathrm{K}$, decreasing to $10\,\mathrm{pJ/K}$ at $1\,\mathrm{K}$, and a low thermal conductance, going from about $3\,\upmu\mathrm{W/K}$ at $300\,\mathrm{K}$ to $8\,\mathrm{nW/K}$ at $1\,\mathrm{K}$. The thermal relaxation time of the device is long enough (ms to s range) to enable ac steady-state and relaxation methods to be used concurrently. The high resolution of the ac steady-state method allows small changes in the heat capacity, such as contributions from the electronic specific heat, to be accurately determined, and investigations of phase transitions and phase diagrams to be performed. Absolute accuracy is obtained by a combination of low background addenda, a stacked calorimeter design, and extensive measurement electronics operated with self-regulation and frequency feedback. The compact format enables the calorimeter to be placed on sample holders for rotation in magnetic field. The calorimeter could even be used for studies of dynamic (frequency dependent) heat capacity.\cite{BIRGE:1985vp} The device is thus a versatile tool for general thermodynamic studies of small samples.

\section{CALORIMETER DESIGN}
\subsection{Fabrication}
\begin{figure}[!tp]
\includegraphics[clip,width=0.95\linewidth]{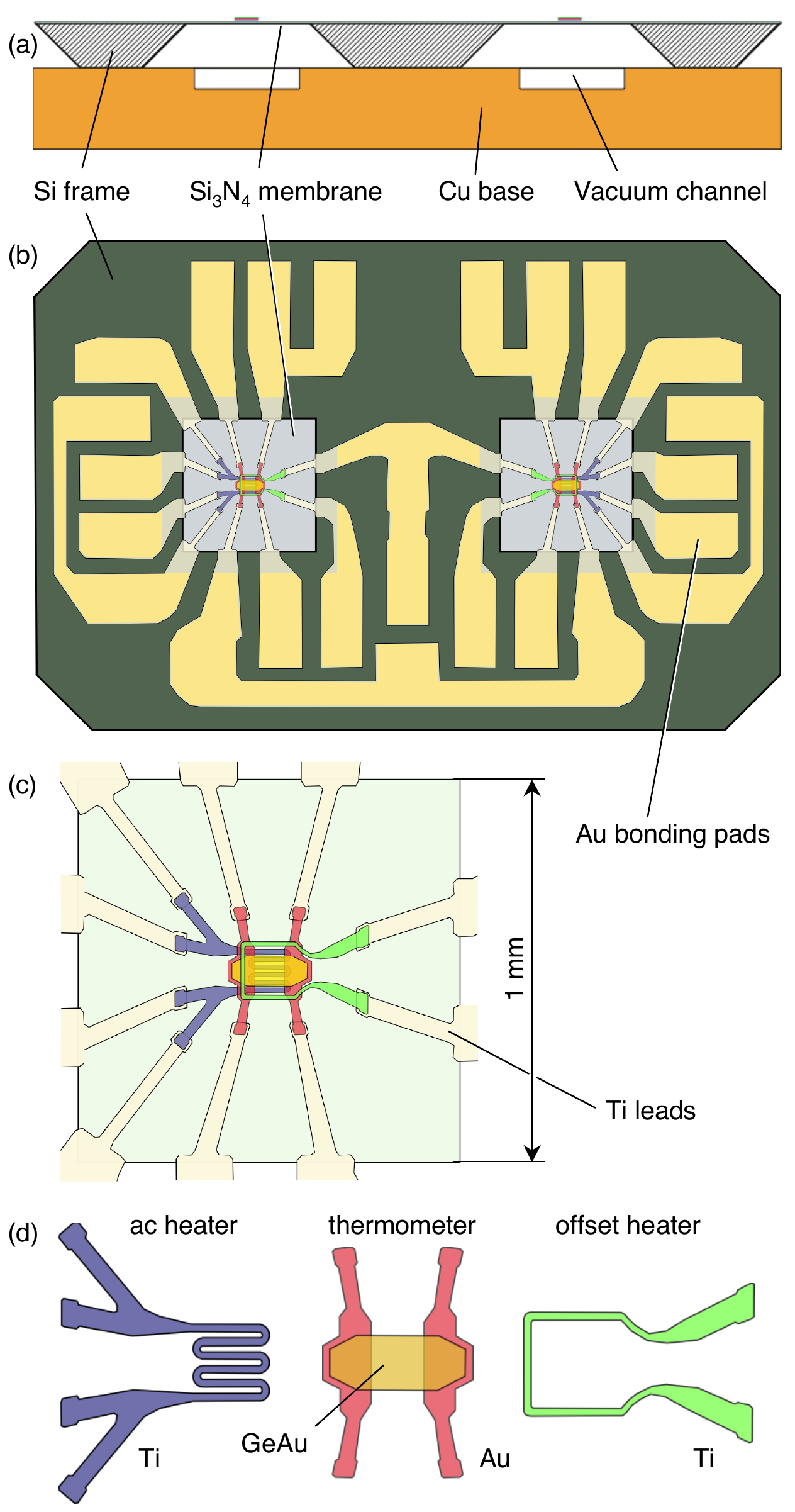}
\caption{\label{Fig1}(a) Cross-sectional schematic of the calorimeter on top of a copper base with vacuum channels. (b) Top view layout of the calorimeter with the two membrane-based calorimeter cells surrounded by 20 bonding pads that connect the calorimeter to the measurement electronics.  (c) Schematic of one of the  $1 \times 1\,\mathrm{mm}^2$ membrane cells, composed by ac heater, thin film GeAu thermometer, and offset heater. The sample is placed on the $80 \times 80\,\upmu\mathrm{m}^2$ central thermometer area. On top of the stack, a thermalization layer made of Au is deposited if smaller samples are to be used, to obtain good internal thermalization and a uniform temperature distribution over the whole thermometer. (d) Illustration of the active layers. All active layers are electrically insulated from each other and the sample by SiO$_2$/AlO$_x$ layers (not shown).}
\end{figure}
The nanocalorimeter is built on top of two custom-designed, pre-fabricated silicon nitride membranes, $1\,\mathrm{mm}\times1\,\mathrm{mm}$ in size and $150\,\mathrm{nm}$ thick (SPI Supplies, West Chester, USA). The membranes are suspended by a Si frame $6.2\,\mathrm{mm}\times3.4\,\mathrm{mm}$, which is attached by means of Stycast to a copper base as illustrated in Fig.~\ref{Fig1}(a). Figure \ref{Fig1}(b) shows a top view of the calorimeter layout. Each cell, shown in Fig.~\ref{Fig1}(c), is composed of a stack of ac heater, thermometer, and offset heater in the central area of the membrane. Between each of these active layers there are electrical insulation layers. All layers are fabricated using photolithography, deposition (e-beam evaporation or sputtering), and double-layer resist lift-off. The active layers are illustrated separately in Fig.~\ref{Fig1}(d).

The ac heater, shown in Fig.~\ref{Fig1}(d), is a meander-shaped resistor made of e-beam evaporated titanium, $10\,\upmu\mathrm{m}$ wide and $50\,\mathrm{nm}$ thick, which covers the central sample area. It is used to oscillate the sample temperature with a well-defined ac power. It has a four-point probe geometry to allow accurate determination of the power at the sample without contributions from lead and contact resistances. Ti becomes superconducting below about $0.4\,\mathrm{K}$, but since the superconductivity can be suppressed by relatively modest fields the heaters may still be used at even lower temperatures. If needed, the film thickness could also be decreased to suppress $T_\mathrm{c}$.

The active part of the thermometer, shown in Fig.~\ref{Fig1}(d), is a $80\,\upmu\mathrm{m} \times 80\,\upmu\mathrm{m}$ square made of $100\,\mathrm{nm}$ thick, sputtered $\mathrm{Ge}_{1-x}\mathrm{Au}_{x}$ alloy. It senses the sample area using a four-point probe configuration and is in direct thermal contact with the sample, thus probing the actual sample temperature. Consequently, heat dissipation in the thermometer does not pose a problem after initial calibration so that the thermometer can be operated at rather high powers to achieve a high sensitivity. Furthermore, there is no need to rely on measurements of the frame/base temperature, unlike when thermocouples are used.\cite{Rydh:2006ta,Graebner:1989jw} This eliminates hysteretic effects when sweeping temperature and results in a high reproducibility. The sensor layer is fabricated using RF magnetron sputtering from a 2-inch target made of cast $\mathrm{Ge}_{1-x}\mathrm{Au}_{x}$ with nominally 17 at\% Au.\cite{Bethoux:1995tz} The chip is annealed at $190^{\circ}\mathrm{C}$ on a hotplate for at least 1 hour after deposition, resulting in a room temperature resistivity $\rho_{RT}\approx 9\,\mathrm{m}\Omega\mathrm{cm}$ and dimensionless sensitivity $\eta = \left| \mathrm{dln}R/\mathrm{dln}T\right| \approx 1$ between $300\,\mathrm{K}$ and $10\,\mathrm{K}$,  increasing to about $2$ at lower temperatures. The GeAu sensor layer is deposited on top of Au leads as shown in Fig.~\ref{Fig1}(d). The leads are in turn connected to external leads in Ti, as seen in Fig.~\ref{Fig1}(c). By combining metals with high (Au) and low (Ti) thermal conductivity a more well-defined isothermal area in the center of the membrane is obtained.\cite{Tagliati:2010jt}

The offset heater, shown in Fig.~\ref{Fig1}(d), is driven by a dc current and can locally increase the sample temperature up to at least $100\,\mathrm{K}$ above the base temperature. This heater is designed to give an isothermal interior area, but is not in direct contact with the active part of the thermometer. It is thus less suitable as an ac heater, since the thermal diffusion between thermometer and heater would have to be taken into account. Titanium has good robustness so the deposited layer can be quite thin. Its addition to the background heat capacity is rather insignificant, but to simplify the fabrication process the layer is sometimes skipped.

The leads that connect the active layers to the external bonding pads are also in Ti. Because of the relatively low thermal conductance of Ti compared to many other metals, the use of Ti for the leads ensures a long relaxation time dominated by the membrane itself. Outside the membrane area, a thick layer of Au is deposited onto the leads to minimize lead resistance effects and to provide a suitable layer for bonding, see Fig.~\ref{Fig1}(a). As a last element of the calorimeter stack, a $110\,\upmu\mathrm{m} \times 110\,\upmu\mathrm{m}$ square may be deposited as a thermalization layer to distribute the temperature evenly over the sample area if small samples are to be studied. This layer may be deposited simultaneously with the outer Au bonding pad layer.

The electrical insulation layers, not shown in Fig.~\ref{Fig1}, are made by thermally evaporated aluminum oxide in combination with sputtered SiO$_2$. The SiO$_2$ is deposited directly following the deposition of heaters or GeAu layer, while the AlO$_x$ layers are patterned as separate layers. There are in total three AlO$_x$ layers in the shape of rounded squares of different sizes: those between heaters and thermometer measure $270\,\upmu\mathrm{m} \times 230\,\upmu\mathrm{m}$ while the last insulation between thermometer and thermalization layer is $180\,\upmu\mathrm{m} \times 180\,\upmu\mathrm{m}$. Care was taken to design the layers so that edges of different layers do not coincide. This greatly reduces the risk of shorts between layers, especially for thin insulation layers.

\begin{figure}
\includegraphics[clip,width=0.95\linewidth]{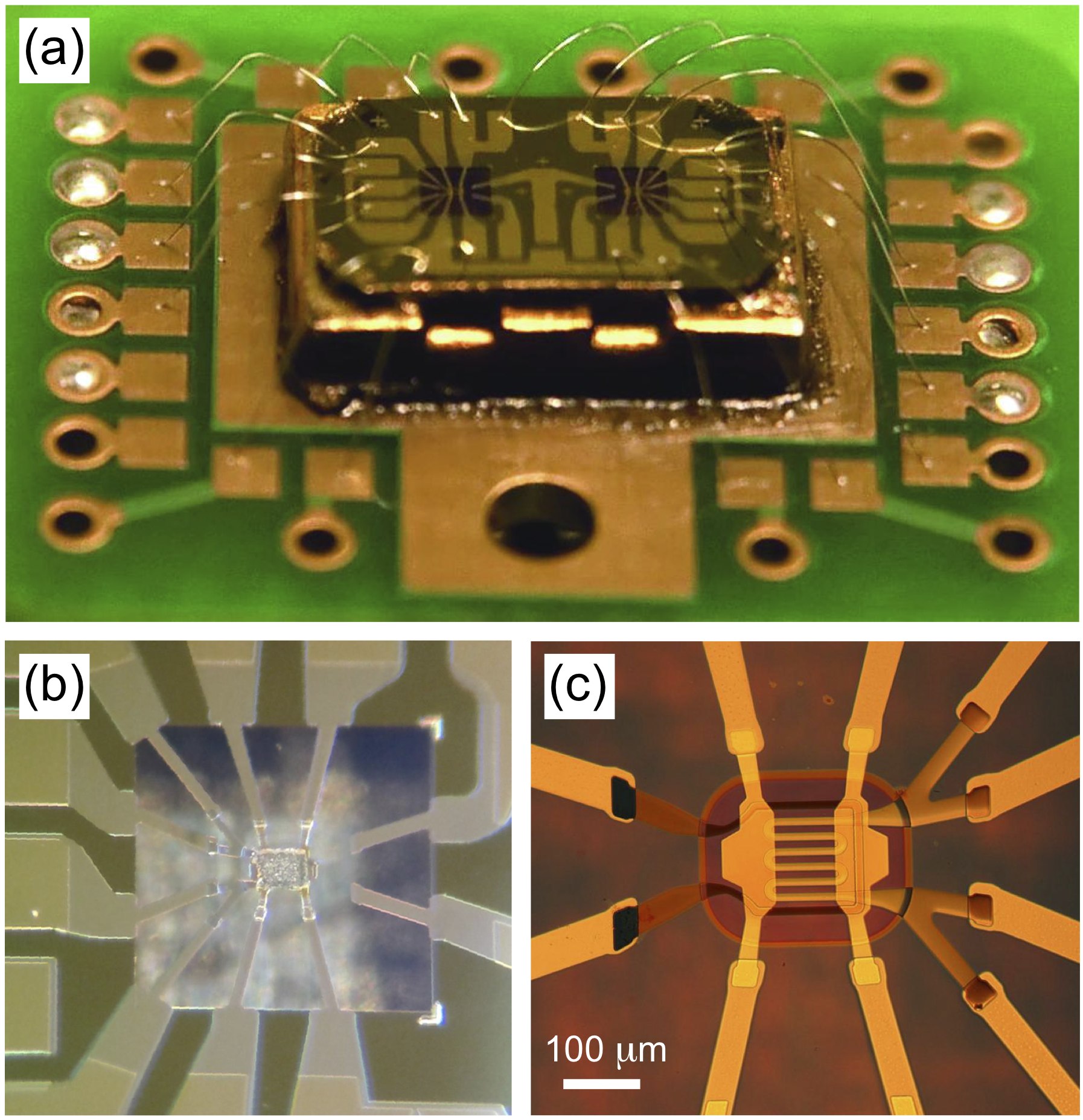}
\caption{\label{Fig2} (a) Calorimeter bonded onto cryostat plug-in. (b) Calorimeter cell with a sample covering ac heater and thermometer. The membrane and electrical insulation layers appear transparent. This particular calorimeter has no offset heater. (c) Microscope image of the central part of a calorimeter cell, illustrating the active layers and insulation (but without GeAu layer for clarity).}
\end{figure}

Figure~\ref{Fig2}(a) shows a picture of the calorimeter bonded onto a cryostat sample holder plug-in. The calorimeter requires up to 20 wires, but can otherwise be fitted onto most cryogenic sample holders. The sample is placed on one of the membrane cells by means of a simple micro-manipulator, or, with some practice, by hand. The other cell may either carry a reference sample or be left empty. Figure~\ref{Fig2}(b) shows a calorimeter cell with a typical sample (the Pb sample discussed in Section~\ref{Sec:Meas}). The central parts of the calorimeter are shown in Fig.~\ref{Fig2}(c), where also the larger electrical insulation layers can be seen.

\subsection{Measurement electronics}
In the standard measurement mode, known ac and dc currents flow through the heaters and thermometers. To practically enable the measurements of temperatures and oscillation amplitudes, a set of time and phase synchronized lock-in amplifiers based on a field-programmable gate array (FPGA) is used.\cite{Rydh:2009wh} We implemented such an instrument using the PXI-7854R card by National Instruments. The card has integrated $750\,\mathrm{kS/s}$ ADCs and $1\,\mathrm{MS/s}$ DACs, providing eight integrated, simultaneous-sampling analog inputs and outputs for ac and dc biasing. For each input, the first and/or second harmonic amplitudes with corresponding phases are extracted, as well as the dc component. A central phase generator delivers the digital reference for all inputs and outputs. The resulting instrument thus allows tuning of output voltages and working frequency during the measurements, without any loss of correlation between inputs and outputs. Each signal, before being read, passes through a low-noise, custom-built preamplifier stage. In total eleven preamplifiers are used, eight of which have variable gain (between 1 and 5000), controlled by the FPGA lock-in, the others with fixed gain (100 or 1000). While measuring, the eight variable-gain preamplifiers are automatically adjusted to maximize their performance. 

\subsection{Thermometer operation}
\begin{figure}
\includegraphics[clip,width=0.95\linewidth]{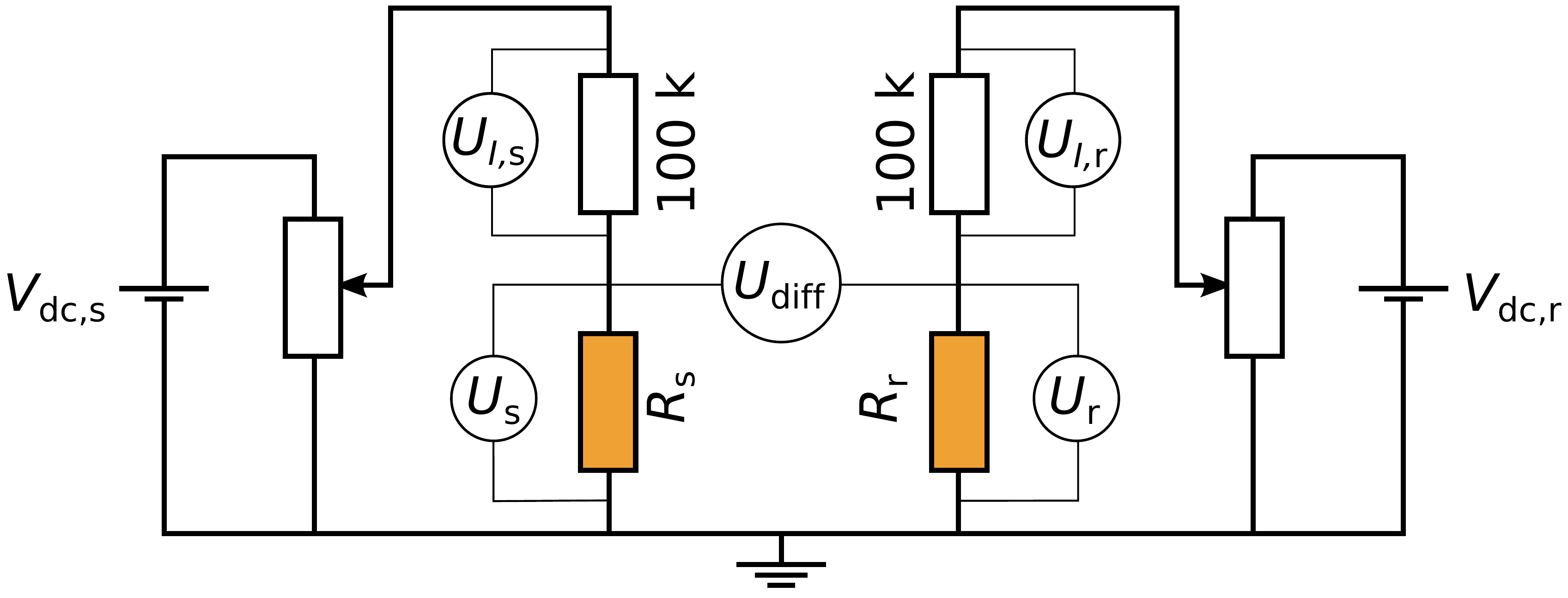}
\caption{\label{Fig3}Thermometer bias and read-out circuit. The sample and reference GeAu thermometer resistances $R_\mathrm{s}$ and $R_\mathrm{r}$ are measured in four-probe configurations with the current provided by synchronous voltage sources $V_\mathrm{dc,s}$ and $V_\mathrm{dc,r}$. The currents through the thermometers are measured by the voltages over $100\,\mathrm{k}\Omega$ series resistors. Optional, adjustable voltage dividers are used to avoid any electrical discharge from destroying the membranes during initial connection.}
\end{figure}

The circuit scheme used for the current bias and voltage read-out of the thermometers is illustrated in Fig.~\ref{Fig3}. An automated process is implemented that auto-adjusts the current through the sample-side thermometer in order to regulate the voltage across the thermometer. The applied voltage $V_\mathrm{dc,s}$ is thus varied to keep the dc component $U_\mathrm{s,dc}$ of the sample voltage $U_\mathrm{s}$ equal to a variable setpoint value that  is typically around $0.1\,\mathrm{V}$ and almost constant over the full temperature range. When the resistance of the thermometer increases with decreasing temperature, the applied current decreases and a suitable power is still delivered. The reading of  $U_\mathrm{s,dc}$ is made synchronously by the lock-in so that frequency-dependent signals are separated from the dc component. Changes to $V_\mathrm{dc,s}$ are synchronized with the measurements as well.

Since the sample and reference thermometers are fabricated together at the same time, they are quite well balanced and display the same temperature dependences, within measurement uncertainty. There may, however, still be a small imbalance in the resistance ratio of the order of 1\% between the two sides, arising from the lithographic tolerance. To compensate for this, a reference adjustment system is applied to vary the reference output $V_\mathrm{dc,r}$ in Fig.~\ref{Fig3} to keep $U_\mathrm{diff,dc}=0$ so that $U_\mathrm{s,dc}= U_\mathrm{r,dc}$, rather than setting the currents or powers equal on the two sides. By balancing the thermometers in this way, the time-varying temperature difference between sample and reference is given by the $U_\mathrm{diff}$ signal through the simple relation
\begin{equation}\label{EqDeltaTemp}
T_\mathrm{diff}=\frac{U_\mathrm{diff} \cdot T}{U_\mathrm{dc} \cdot \mathrm{dln}R/\mathrm{dln}T}.
\end{equation}
Here $T$ is the absolute temperature, which is obtained from the dc component of the thermometer resistance. The additional temperature difference caused by the power difference is usually insignificant, but can be compensated for by a corresponding power from the offset heater. The absence of dc offsets in the bridge differential $U_\mathrm{diff}$ makes it possible to apply a high amplification to $U_\mathrm{diff}$ without overloading, which results in highest possible resolution. The close proximity of sample and reference thermometers further eliminates common noise sources such as electromagnetic interference and temperature variations of the base frame. It should be noted that Eq.~(\ref{EqDeltaTemp}) has to be modified when the thermometer resistance becomes comparable to the series resistance of Fig.~\ref{Fig3}, to include the effect of a varying bias current.

\section{MEASUREMENT TECHNIQUES}
\subsection{AC steady state}
In the ac steady-state method \cite{Sullivan:1968uo} the sample temperature is made to oscillate with an amplitude typically in the range $1$\,--\,$100\,\mathrm{mK}$. This modulation is created by an ac power, which for resistive heating is given by $P(t)=R_\mathrm{h}I^{2}_{0}(1+\sin\omega t)=P_{0}(1+\sin\omega t)$. It is thus generated by an ac current with RMS amplitude $I_{0}$ and angular frequency $\omega/2$ flowing through the ac heater resistor $R_\mathrm{h}$. The temperature response of the cell is given by $T(t)=T_\mathrm{b}+T_\mathrm{offs}+T_\mathrm{ac}(t)$. $T_\mathrm{b}$ is the base temperature, $T_\mathrm{off}=P_0/K_\mathrm{e}$ is the dc offset due to the time-averaged power supplied by the heater resistance, where $K_\mathrm{e}$ is the thermal conductance between sample and thermal bath (Si frame), and $T_\mathrm{ac}(t)$ is the oscillating term whose steady-state amplitude $T_\mathrm{ac}$ is directly related to the heat capacity of the sample:
\begin{equation}\label{EqTac}
 T_\mathrm{ac}=\frac{P_{0}}{\omega C}\left[1+\frac{1}{(\omega \tau_\mathrm{e})^2}+f(\tau_\mathrm{i})\right]^{-1/2}.
\end{equation}
Here $\tau_\mathrm{e}=C/K_\mathrm{e}$ is the external relaxation time between sample and external thermal bath and $f(\tau_\mathrm{i})$ is a rather complicated function of the internal relaxation time between sample and calorimetric cell, $\tau_\mathrm{i}=C_\mathrm{s}/K_\mathrm{i}$, where $C_\mathrm{s}$ is the sample heat capacity and $K_\mathrm{i}$ is the thermal conductance between the sample and the central cell platform. The value of $K_\mathrm{i}$ strongly depends on the agent used to attach the sample to the nanocalorimetric cell. 

Equation~(\ref{EqTac}) is impractical to use to obtain $C$ from $T_\mathrm{ac}$, due to the complications of $f(\tau_\mathrm{i})$.
We have, however, previously shown\cite{Tagliati:2011be} that for a system with good thermal connection between the active layers, but with possibly significant $\tau_\mathrm{i}$, the thermometer temperature oscillation $T_\mathrm{ac,0}$ and corresponding phase $\phi$ between power and temperature oscillation can be found as
\begin{numcases}{}\label{EqTPhi}
\begin{aligned}
T_\mathrm{ac,0} &=\frac{P_\mathrm{0}}{\sqrt{(\omega C)^{2}+K^{2}}}	\\
\tan\phi &=\frac{\omega C}{K} 			
\end{aligned}
\end{numcases}
provided that $C$ and $K$ are taken as
\begin{numcases}{}
\begin{aligned}
C &= C_\mathrm{cell} + (1-g) C_\mathrm{s}\\
K &=K_\mathrm{e,eff}+g K_\mathrm{i}\\
g &= \frac{(\omega \tau_\mathrm{i})^2}{1+(\omega \tau_\mathrm{i})^2}
\end{aligned}\label{Eq_CKtotal}
\end{numcases}
Here, $C_\mathrm{cell}$ is the empty cell contribution to the heat capacity and $K_\mathrm{e,eff}$ is the effective thermal link of the membrane.\cite{Tagliati:2011be} Equation~(\ref{EqTPhi}) can be wrapped around into the practical, functional relations
\begin{numcases}{}\label{EqCK}
\begin{aligned}
 C &=\frac{P_\mathrm{0}}{\omega T_\mathrm{ac,0}}\sin\phi		\\
 K &=\frac{P_\mathrm{0}}{T_\mathrm{ac,0}}\cos\phi 			
\end{aligned}
\end{numcases}
which form the basis of evaluating the measurements. Note that the phase is carrying information that is needed to achieve good absolute accuracy. The phase can be used to verify that the frequency is selected correctly, i.e., that $\omega\tau_\mathrm{i}\ll1$ so that $g \to 0$. If the frequency is so high that $\sin \phi \to 1$ in Eq.~(\ref{EqCK}), it is very likely that $g$ contributes significantly in Eq.~(\ref{Eq_CKtotal}). A known phase is also necessary to accurately extract $K_\mathrm{e}$ from ac steady-state measurements.

When using the ac steady-state technique in practical terms, we apply an ac current to the heater(s) and measure the thermometer and series resistance voltages $U_\mathrm{s}$, $U_\mathrm{r}$, $U_{I,\mathrm{s}}$, $U_{I,\mathrm{r}}$, and $U_\mathrm{diff}$, defined in Fig.~\ref{Fig3}, at the second harmonic of the heater current frequency (simultaneously with the synchronous dc mean). The amplitudes of the sample and reference temperature oscillations $T_\mathrm{s,ac}$ and $T_\mathrm{r,ac}$, are related to the corresponding measured voltages over the thermometers and currents through the reference resistors by 
\begin{equation}\label{TacFromUac}
T_{\mathrm{ac}}=\left({\frac{U_\mathrm{ac}}{U_\mathrm{dc}}+\frac{I_\mathrm{ac}}{I_\mathrm{dc}}}\right)\frac{T}{\eta}.
\end{equation}
Here both $T$ and sensitivity $\eta$ are obtained from the dc measurement of the thermometer resistance $R$ and previous calibration of $R(T)$.  For the temperature differential the corresponding expression is
\begin{equation}\label{TdiffFromUdiff}
T_\mathrm{diff,ac}=\frac{U_\mathrm{diff,ac}}{U_\mathrm{dc}}\left({1+\frac{U_\mathrm{dc}}{U_{I,\mathrm{dc}}}}\right)\frac{T}{\eta},
\end{equation}
where $U_{I,\mathrm{dc}} = R_\mathrm{ref}I_\mathrm{dc}$ and $R_\mathrm{ref}$ is the series resistance for current measurement. All ac signals in Eqs.~(\ref{TacFromUac}) and (\ref{TdiffFromUdiff}) refer to amplitudes in the steady state. Note that $T_\mathrm{diff,ac} \ne T_\mathrm{s,ac} - T_\mathrm{r,ac}$ in general, since the sample and reference temperature oscillations may have different phases.

In the simplest measurement case, the reference heater power is kept off. This method is used when the heat capacity of the sample is fairly large as compared to the heat capacity of the empty cell and no reference sample is placed on the reference side. The sample temperature oscillation is then measured through the differential signal $U_\mathrm{diff,ac}=U_\mathrm{s,ac}$ while $U_\mathrm{r,ac}=0$. Since $U_\mathrm{diff}$ has no dc offset, the differential signal yields a significantly higher resolution than $U_\mathrm{s,ac}$. To subtract $C_\mathrm{cell}$ from $C$, sample and empty cell are measured in separate runs. The measurement of $T_\mathrm{ac}$ is thus made differentially, but the measurement of $C$ is not.

If the sample is small, or if a similar-size reference sample is added to the reference side, a truly differential measurement mode can be employed. In this case, the same power is applied to both sample and reference sides ($P_\mathrm{s}=P_\mathrm{r}=P_\mathrm{0}$). The heat capacities $C_\mathrm{s+cell}$ and $C_\mathrm{r+cell}$ of the individual sides are still given by Eq.~(\ref{EqCK}), but the differential heat capacity $C_\mathrm{diff}=C_\mathrm{s}-C_\mathrm{r}$ can be obtained as well from $T_\mathrm{diff,ac}$:
\begin{equation} \label{C_diff}
C_\mathrm{diff}=\frac{P_0 T_\mathrm{diff,ac}}{\omega T_\mathrm{s,ac} T_\mathrm{r,ac}}.
\end{equation}
Equation~(\ref{C_diff}) is valid under the assumption that $K$ is the same for both sample and reference sides. Since $K_\mathrm{e}$ is given mainly by the membrane itself, and other contributing layers are manufactured lithographically, this requirement is easy to fulfill in normal cases. If a large sample is studied, the sample itself may enhance $K$ through a small, but nonzero $g$. The absolute accuracy of Eq.~(\ref{C_diff}) may in this case be tested by a separate verification measurement with $P_\mathrm{r}=0$.

If the differential heat capacity mode is used with a large sample but without reference, the signal from the reference side will dominate $U_\mathrm{diff}$, and the benefit of a differential measurement would be lost. An alternative to reverting to a single-side measurement can in this case be to adjust the amplitude and output phase of the reference heater power to maintain $U_\mathrm{diff}=0$. This mode of measurement is, however, still unexplored.

\subsection{Frequency feedback}
\begin{figure}
\includegraphics[clip,width=0.95\linewidth]{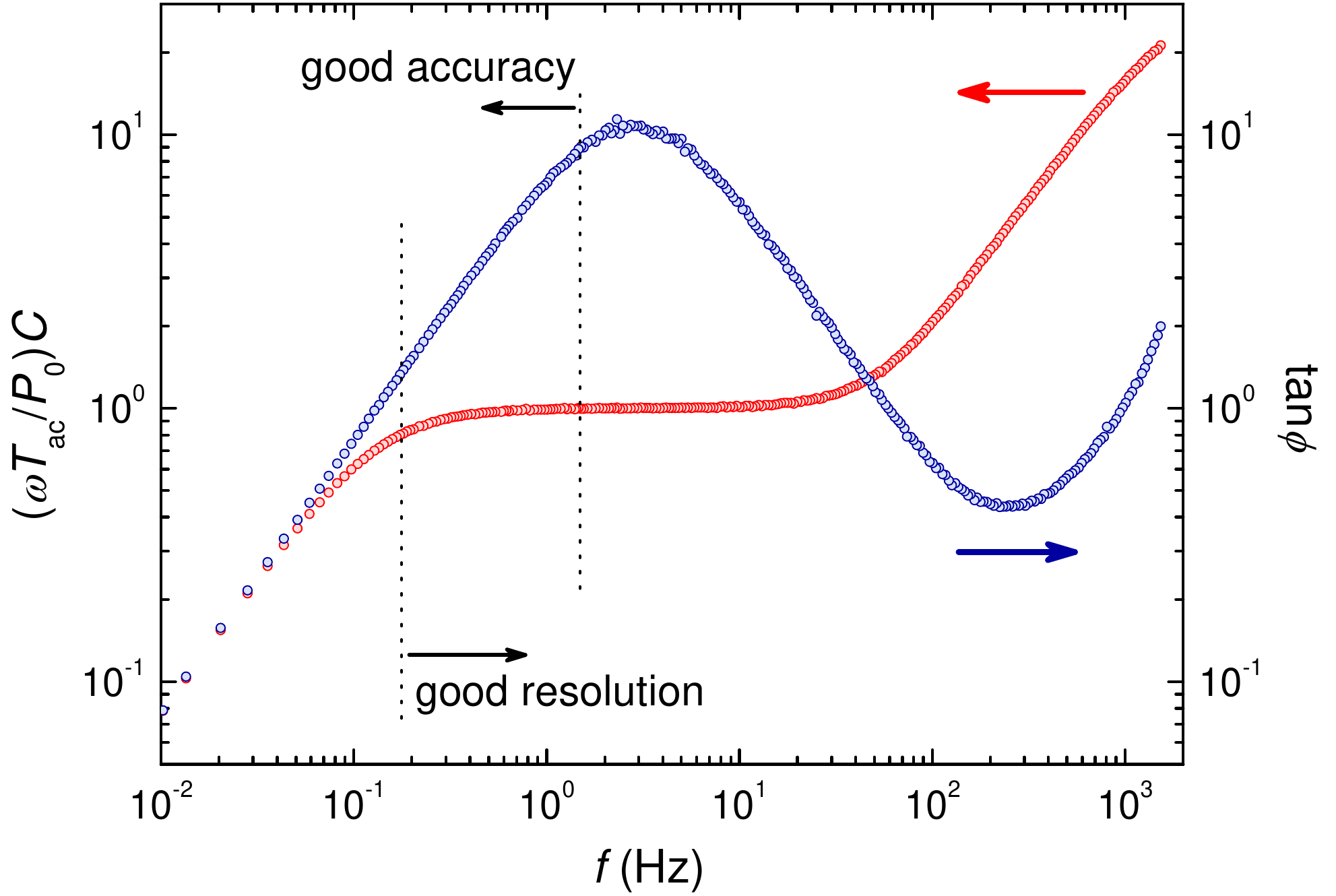}
\caption{\label{Fig4}Measured frequency dependence of temperature oscillation amplitude $T_\mathrm{ac}$ and phase $\phi$, expressed as $(\omega T_\mathrm{ac}/P_\mathrm{0})C$ and $\tan \phi$, respectively, with $P_\mathrm{0}$ and $C$ constant ($f=\omega/4\pi$). At low frequency $\tan \phi \sim \omega$ and measurements yield good absolute accuracy. At too low frequencies, however, the signal no longer comes from the heat capacity but from the thermal link, and the resolution decreases. Note that the middle of the adiabatic plateau (where $\omega T_\mathrm{ac}$ is constant)  is not corresponding to the best measurement frequency if good accuracy is required.}
\end{figure}

We have previously shown\cite{Tagliati:2011be} that good absolute accuracy can be obtained provided that one has good control of the phase in Eq.~(\ref{EqCK}). In Fig.~\ref{Fig4}, the frequency dependence of $T_\mathrm{s,ac}$ and $\phi$ is shown for a typical sample.
Good accuracy is found at low frequencies, where $\tan \phi \sim \omega$. At higher frequencies, the effect of $\omega\tau_\mathrm{i}$ is no longer negligible and the accuracy is quickly deteriorating. For the resolution, the conditions are the opposite; at high frequencies the resolution is good, but at low frequencies, $\omega\tau_\mathrm{e}\ll 1$, Eq.~(\ref{EqTPhi}) is reduced to $T_\mathrm{ac}=P_0/K_\mathrm{e}$ and heat capacity is no longer probed.

By fixing the phase $\phi$ to a constant value in the range where measurements yield both good absolute accuracy and good resolution, optimal conditions are found. We do this by continuously adjusting the frequency $\omega$ during the measurement to keep $\tan \phi$ constant by means of an auto-tuning routine in the FPGA lock-in. By having a control loop time equal to a single sample of the ADC, the frequency is smoothly adjusted without introducing additional noise to the measurements. The frequency is thus varying even during a single cycle of the output. This is possible thanks to the synchronous sampling. It should be noted that the middle of the adiabatic plateau (with $\omega T_\mathrm{ac} \sim \mathrm{const.}$) is typically at too high frequencies for good absolute accuracy, as seen in Fig.~\ref{Fig4}.

\subsection{Thermal relaxation}
The thermal relaxation method \cite{Bachmann:1972vh} consists of applying a known power to the heater to raise the sample temperature an amount $\Delta T$ above the base temperature $T_\mathrm{b}$. After a stable sample temperature has been reached, the heater power is turned off and the temperature is allowed to relax back to $T_\mathrm{b}$. The time dependence of the relaxation is exponential and depends on the external time constant of the system:
\begin{equation}
T(t)=T_\mathrm{b}+\Delta T e^{-t/\tau_\mathrm{e}}.
\end{equation}
When using the thermal relaxation method, we apply a square wave with sufficiently low repetition rate to the sample heater, while keeping the reference heater off. The induced temperature response is directly given by Eq.~(\ref{EqDeltaTemp}), from which $\Delta T$ and $\tau_\mathrm{e}$ can be obtained.
The thermal conductance $K_\mathrm{e}$ between sample and base frame is then found as 
\begin{equation} \label{Ke_relax}
K_\mathrm{e}=\frac{\Delta P}{\Delta T},
\end{equation}
where $\Delta P$ is given by the directly measured step in heater power. The sample heat capacity including cell addenda is finally given by $\tau_\mathrm{e}$ and $K_\mathrm{e}$, according to
\begin{equation} \label{C relax}
C_\mathrm{s}+C_{\mathrm{cell}}=\tau_\mathrm{e} \cdot K_\mathrm{e}.	
\end{equation}
To practically perform relaxation measurements, a routine was developed that automatically acquires the voltage pulses, fits the exponential decay to data within 10\% and 90\% of the relaxation, and determines $\Delta T$. In this way, the heat capacity and device thermal conductance can be measured as a function of temperature or magnetic field with immediate results displayed on the screen.

\section{CALIBRATION AND CHARACTERIZATION}
\subsection{Thermometer calibration}

\begin{figure}
\includegraphics[clip,width=0.95\linewidth]{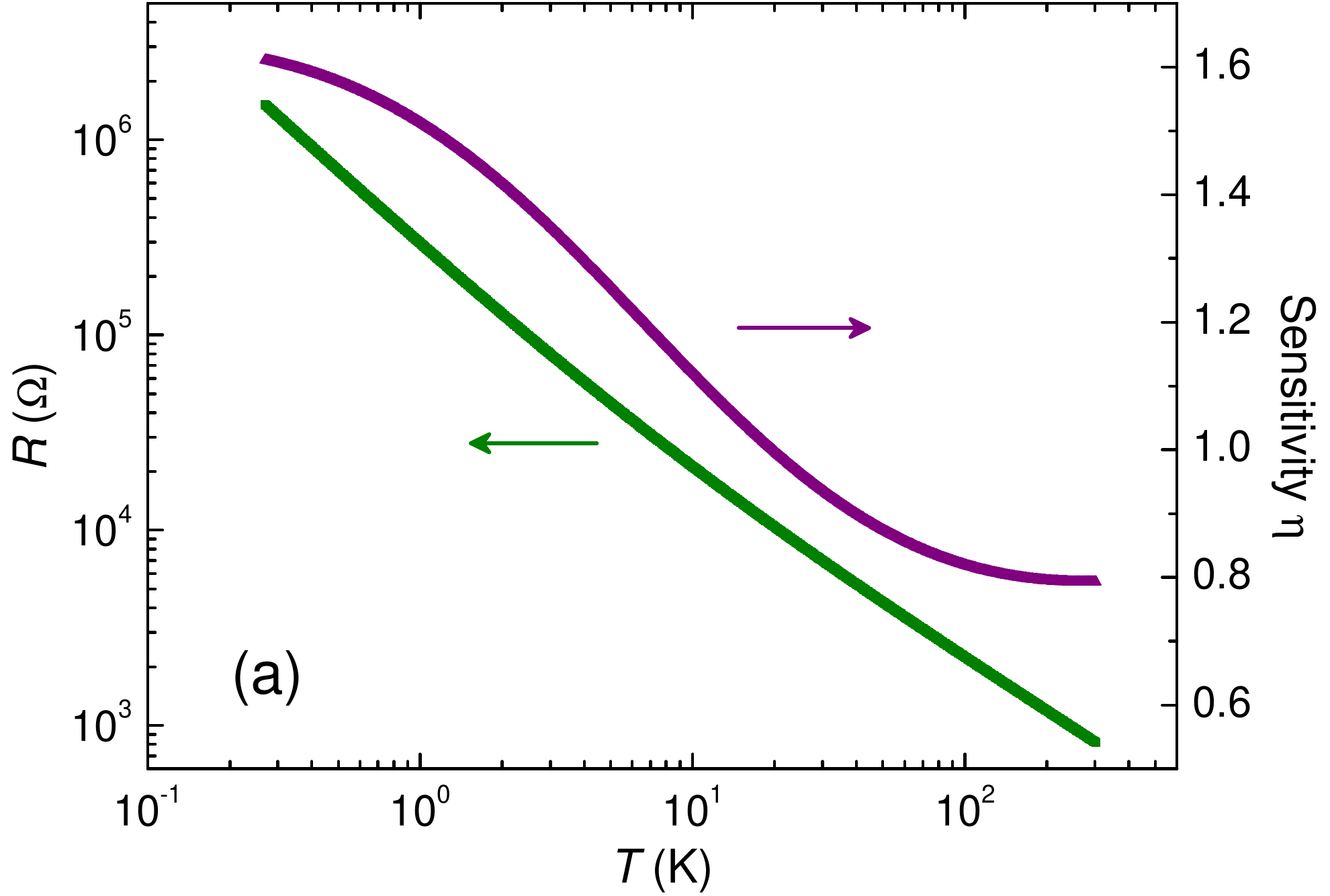}\\[1mm]
\includegraphics[clip,width=0.95\linewidth]{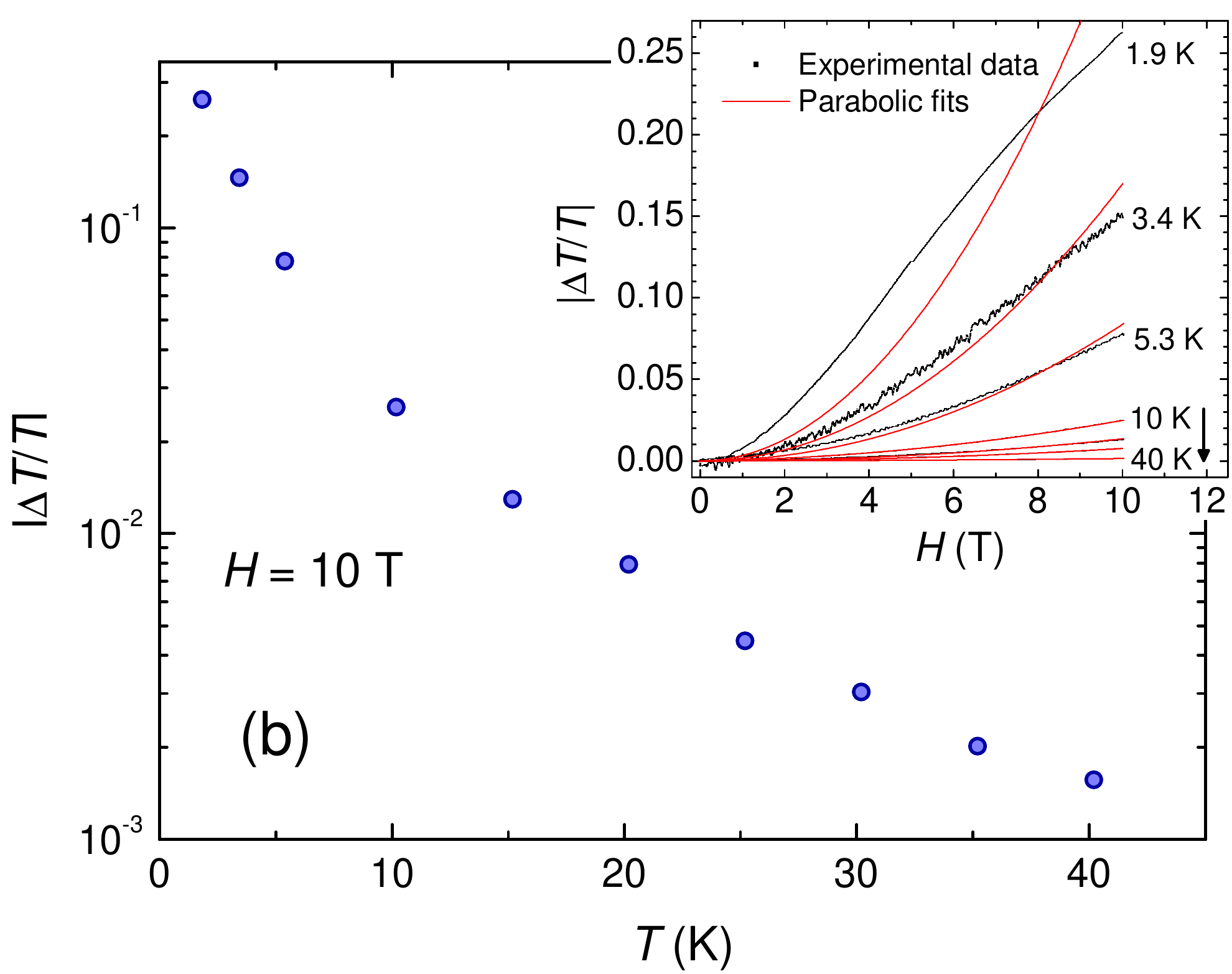}
\caption{\label{Fig5}(a) Temperature dependence of the GeAu thermometer resistance and dimensionless sensitivity $\eta=\left| \mathrm{dln}R/\mathrm{dln}T \right|$. (b) Thermometer magnetoresistance expressed as apparent relative temperature change in a magnetic field of $H=10\,\mathrm{T}$. The inset shows the corresponding field dependence at different temperatures. The magnetoresistance is following a parabolic field dependence except for temperatures below $\sim 3\,\mathrm{K}$ at high fields.}
\end{figure}

Before any measurements, the GeAu thermometers on sample and reference sides need to be calibrated against a known thermometer. We use a Cernox sensor on the sample holder. The calibration is done in the presence of a few mbar helium gas to increase the thermal link between GeAu thermometers and Si base frame, which quickly reduces $\tau_\mathrm{e}$ and temperature offsets due to self-heating. The typical temperature dependence of the GeAu thermometer resistance is shown in Fig.~\ref{Fig5}(a). Once the calibration curve has been obtained, one can attain the temperature dependence of the sensitivity $\eta$ which is approximately constant from $300\,\mathrm{K}$ down to about $50\,\mathrm{K}$, below which it increases slowly as shown in Fig.~\ref{Fig5}(a). The temperature dependence of the sample to reference thermometer resistance ratio $R_\mathrm{s}(T)/R_\mathrm{r}(T)$ is constant within experimental uncertainty over the entire temperature range. The same calibration curve can thus be used for both sides, except for a scaling pre-factor. Thermometers deposited in the same sputtering cycle display very similar temperature dependences, while thermometers fabricated at different times require individual calibrations, although differences are small enough for a standard calibration curve to be used initially in most cases. The thermometer resistance is stable over time provided that the thermometer is not heated excessively after the heat treatment at $190^\circ\mathrm{C}$.

To obtain $\eta$ and $T$ from $R$ while avoiding numerical noise and other artifacts, we fit an analytical expression to the calibration data. The thermometer  $T(R)$ relation is well described by
\begin{equation} \label{Tfit}
T = \frac{T_0}{(R/R_0)^{\alpha_0}}+\frac{T_1}{(R/R_1)^{\alpha_1}}.	
\end{equation}
Here $R_i$, $\alpha_i$, and $T_i$ are constants. Remaining small deviations are fitted by a high-degree polynomial in $\log{(R/R_2)}$. An analytical expression of $\eta$ is then directly found from the fit parameters, and is given by
\begin{equation} \label{Etafit}
1/\eta = \left| \frac{\mathrm{dln}T}{\mathrm{dln}R} \right| = \frac{1}{T} \left[ \frac{\alpha_{0}T_0}{(R/R_0)^{\alpha_0}}+\frac{\alpha_{1}T_1}{(R/R_1)^{\alpha_1}} \right],
\end{equation}
up to the contribution from the polynomial correction. This procedure to obtain temperature and sensitivity from resistance avoids interpolations, which, even in log-log scale, tend to give rise to artificial kinks that may become significant in certain cases such as when studying specific heat differences.

\subsection{Thermometer magnetoresistance}
Resistive thermometers require corrections for magnetic-field-induced changes at low temperature. The magnetoresistance of the GeAu sensor was studied by sweeping a magnetic field between $-10\,\mathrm{T}$ and $10\,\mathrm{T}$ at several fixed temperatures from $1.9\,\mathrm{K}$ to $40\,\mathrm{K}$. The relative change in apparent temperature was then calculated as $\Delta T/T=(\Delta R/R)/\eta$, where $\eta$ is the zero field sensitivity. The curves, shown in the inset of Fig.~\ref{Fig5}(b), can be fairly well approximated by a parabolic field dependence $\Delta R \propto H^2$ above $3\,\mathrm{K}$. The magnetoresistance is positive and its magnitude decreases quite quickly with increasing temperature. The field-induced error at $10\,\mathrm{T}$, if no correction is made, is $1\%$ at $17\,\mathrm{K}$ and $0.15\%$ at $40\,\mathrm{K}$, as shown in Fig.~\ref{Fig5}(b).

\subsection{Empty cell characterization}
\begin{figure}
\includegraphics[clip,width=0.95\linewidth]{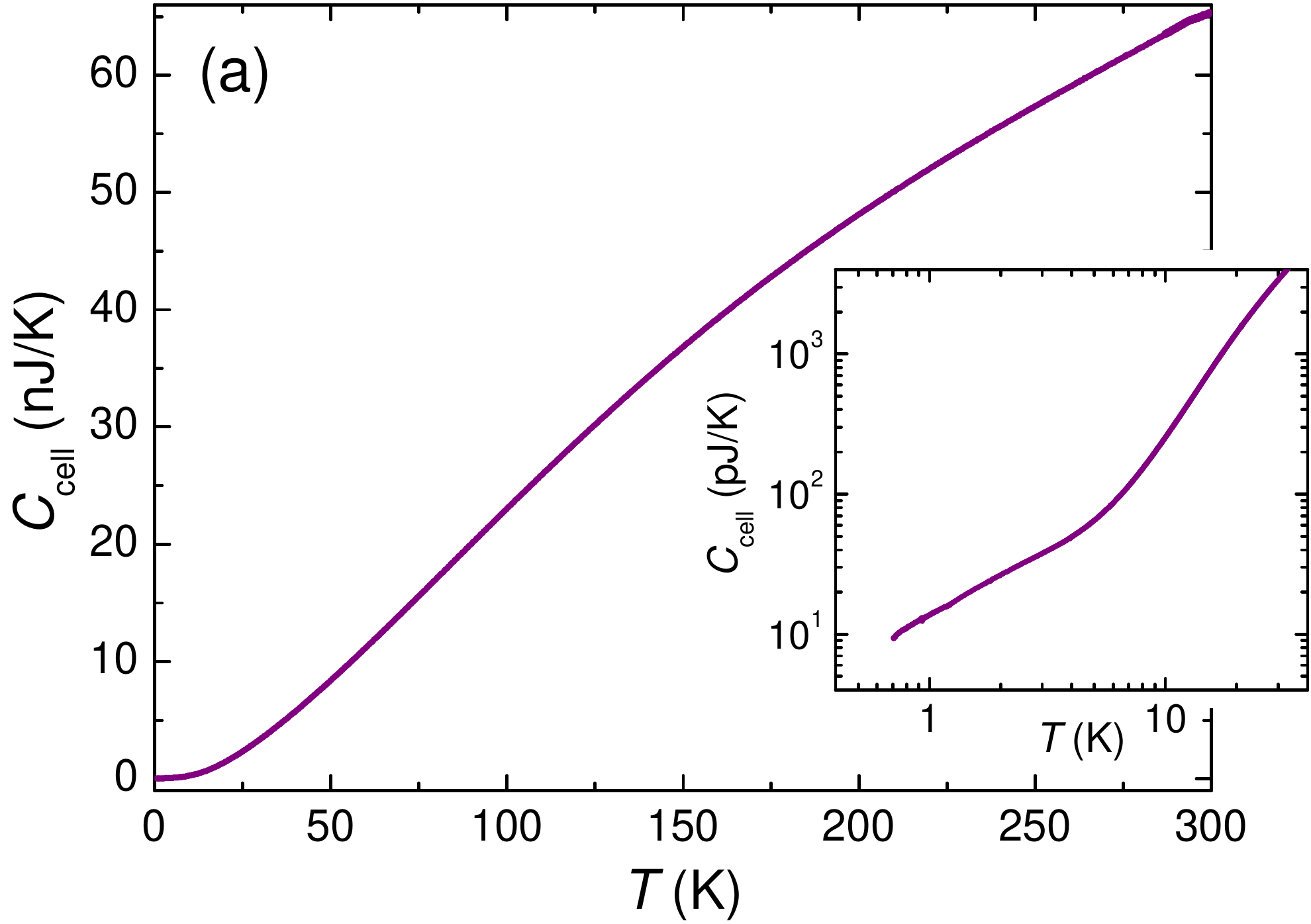}\\[1mm]
\includegraphics[clip,width=0.95\linewidth]{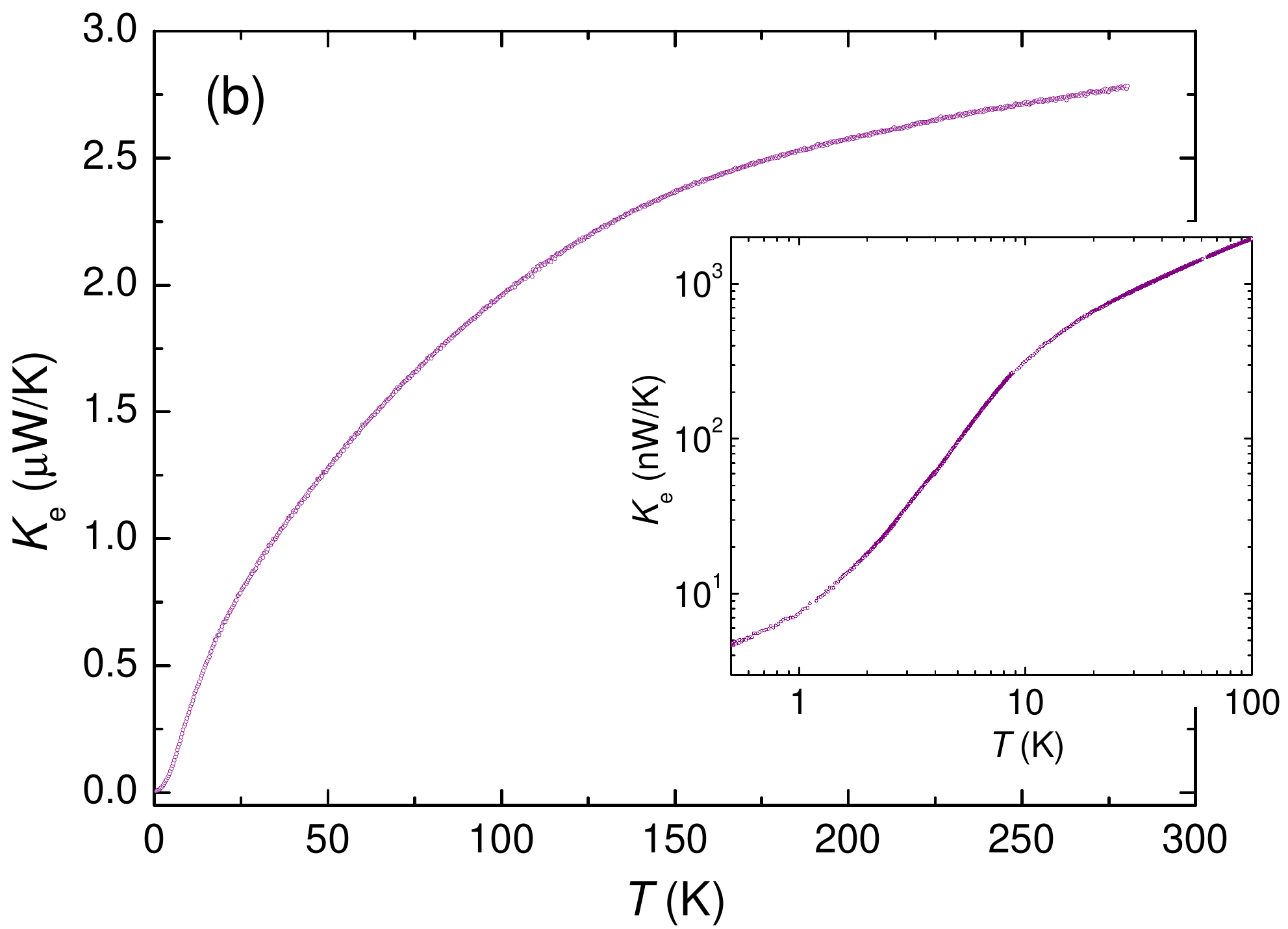}
\caption{\label{Fig6}(a) Empty cell heat capacity $C_{\mathrm{cell}}$ as a function of temperature. (b) Device thermal conductance $K_\mathrm{e}$. The insets show the low-temperature behavior in log-log scale. Note that $C_{\mathrm{cell}}$ and $K_\mathrm{e}$ display a fairly similar temperature dependence, resulting in a rather constant time constant $\tau_\mathrm{e,cell}$ as a function of temperature for the empty device.}
\end{figure}

The nanocalorimeter was characterized thoroughly from room temperature down to $0.5\,\mathrm{K}$. Figure~\ref{Fig6}(a) shows the heat capacity of the empty cell as obtained from Eq.~(\ref{EqCK}).
This heat capacity will, under certain conditions, depend on the choice of woking frequency (i.e.\ $\tan \phi$), since the amount of membrane that is temperature-modulated decreases with increasing frequency.\cite{Tagliati:2011be} The characteristic frequency of the membrane is, however, typically higher than the frequency of normal measurements, making it possible to treat $C_{\mathrm{cell}}$ as a reproducible background addenda for a given calorimeter and to assume that $C_\mathrm{cell}$ is frequency independent for large samples (i.e.\ with $C>C_\mathrm{cell}$). In differential mode, the addenda heat capacity is normally less than 5\% of $C_{\mathrm{cell}}$. A typical noise level is $\delta C \sim 2\,\mathrm{pJ/K}$ at $50\,\mathrm{K}$. The noise level expressed as $\delta C/C$ is fairly constant over the entire temperature range. With a typical measurement time of $3\,\mathrm{s}$ and a moderately large $T_\mathrm{ac}/T$, it is possible to reach $\delta C/C \sim 10^{-4}$.

The thermal link $K_\mathrm{e}$ of the calorimeter cell is shown in Fig.~\ref{Fig6}(b). It can be obtained from Eq.~(\ref{EqCK}) provided that the measurements are made at low enough frequency, $\omega \tau_\mathrm{i} \ll 1$, or through relaxation measurements. While the cell heat capacity is dominated by the membrane that has a fairly high Debye temperature, the thermal link is given by a combination of membrane and metallic leads. The characteristic time constant $\tau_\mathrm{e,cell}=C_\mathrm{cell}/K_\mathrm{e}$ is nevertheless fairly temperature independent.

\subsection{Operational parameters}
\begin{figure}
\includegraphics[clip,width=0.95\linewidth]{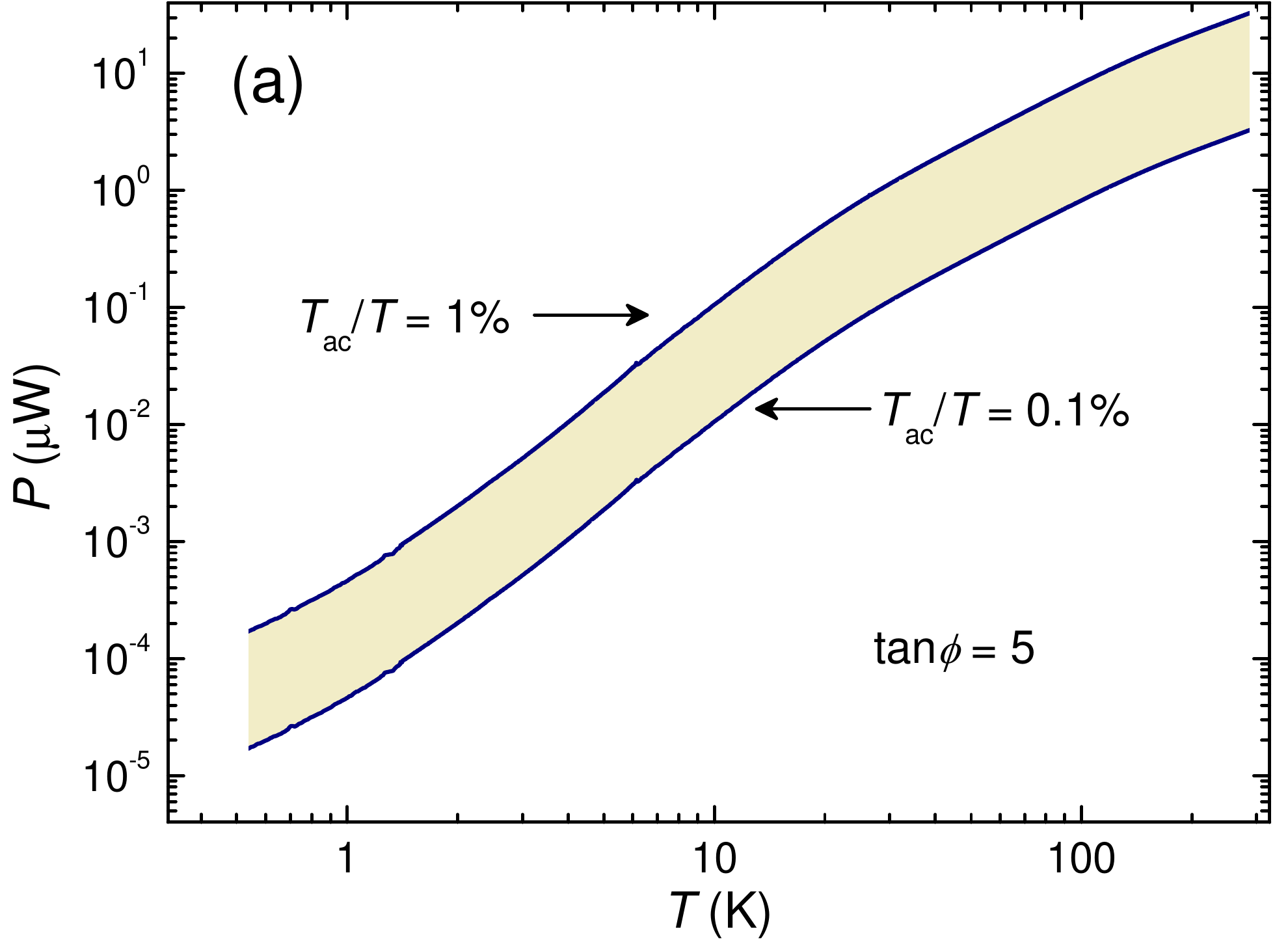}\\[1mm]
\includegraphics[clip,width=0.95\linewidth]{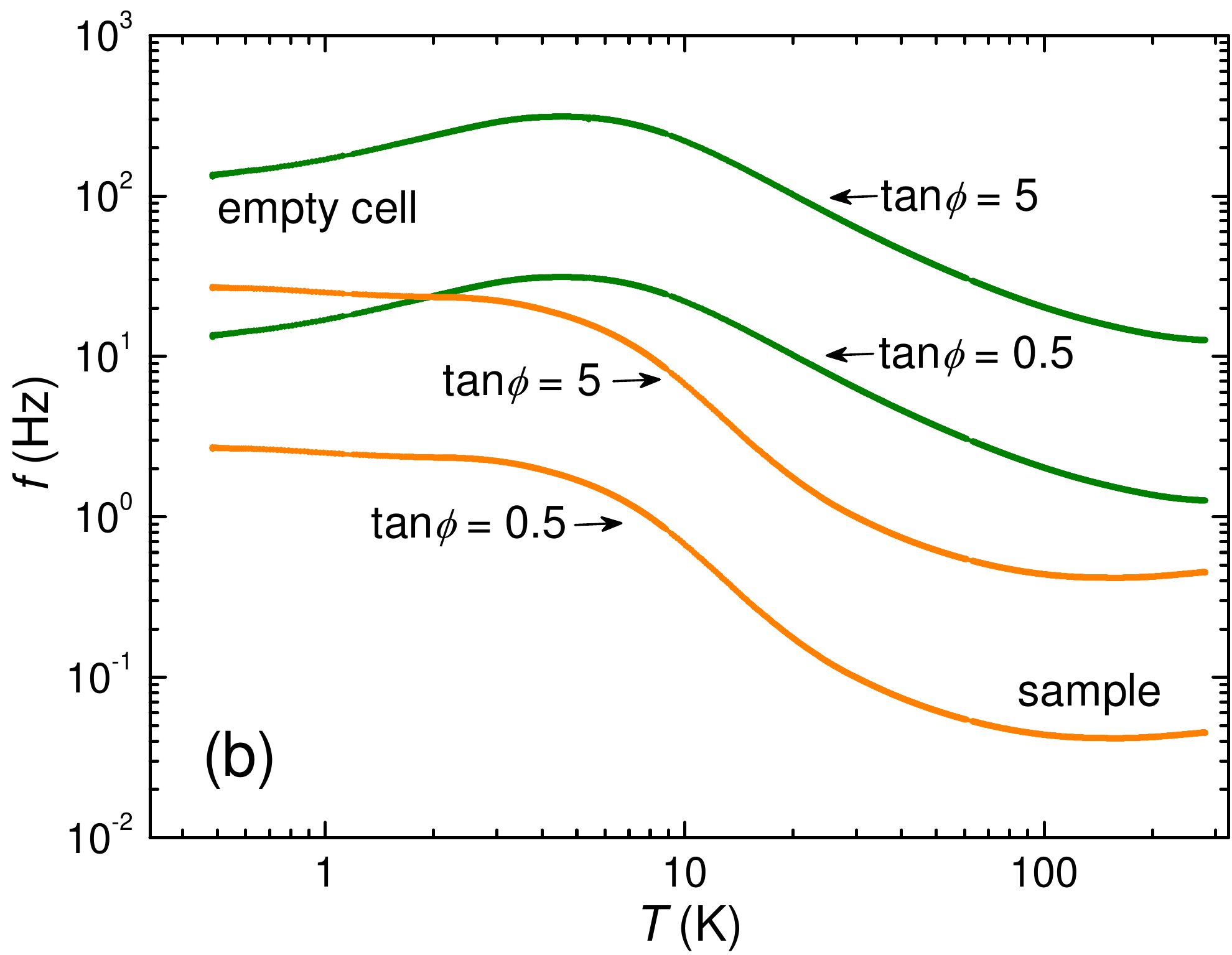}
\caption{\label{Fig7}(a) Range of operational ac heater power as a function of temperature for measurement frequency adjusted so that $\tan \phi = 5$. (b) Typical range of working frequency $f$ as a function of temperature, with and without sample. Note that the sample specific heat controls the frequency range over which good absolute accuracy is obtained. Different samples thus require different frequency adjustment.}
\end{figure}

Figure~\ref{Fig7}(a) shows the range of operational power of the ac heater as a function of temperature. The power required for a certain ratio $T_\mathrm{ac}/T$ between temperature oscillation amplitude and absolute temperature is not depending on $C$ but only $K_\mathrm{e}$, provided that the phase $\phi$ is kept constant. An easy way to adjust the power to the proper level is thus to maintain a constant ratio $U_\mathrm{s,ac}/U_\mathrm{dc} \simeq (T_\mathrm{s,ac}/T)\eta$ while adjusting the frequency so that $\phi$ is constant. In this way, the temperature offset due to the ac heater power is always a constant fraction of $T$ as well. To maintain a similar power for the dc bias of the thermometer, the $U_\mathrm{s,dc}$ should decrease from about $0.1\,\mathrm{V}$ at $100\,\mathrm{K}$ to $0.01\,\mathrm{V}$ at $1\,\mathrm{K}$. In practice, $U_\mathrm{s,dc}$ can be kept almost constant, so that the relative temperature offset due to thermometer self-heating is somewhat higher at the lowest temperatures and somewhat lower at the highest.

Figure~\ref{Fig7}(b) shows the typical range of working frequency $f$ as a function of temperature, with and without sample. For the empty device, the time constant $\tau_\mathrm{e,cell}$ is rather temperature independent, leading to a fairly constant $f$. With a sample, the frequency $f \sim K/C$ is lowered, but the frequency also becomes more strongly temperature dependent. This variation depends on the sample heat capacity. It is thus clear that only narrow temperature ranges can be studied with combined good accuracy and resolution if a constant frequency is used as in the case of traditional ac calorimetry.

\section{MEASUREMENTS}\label{Sec:Meas}
\subsection{Comparing relaxation and ac steady-state methods}
\begin{figure}
\includegraphics[clip,width=0.95\linewidth]{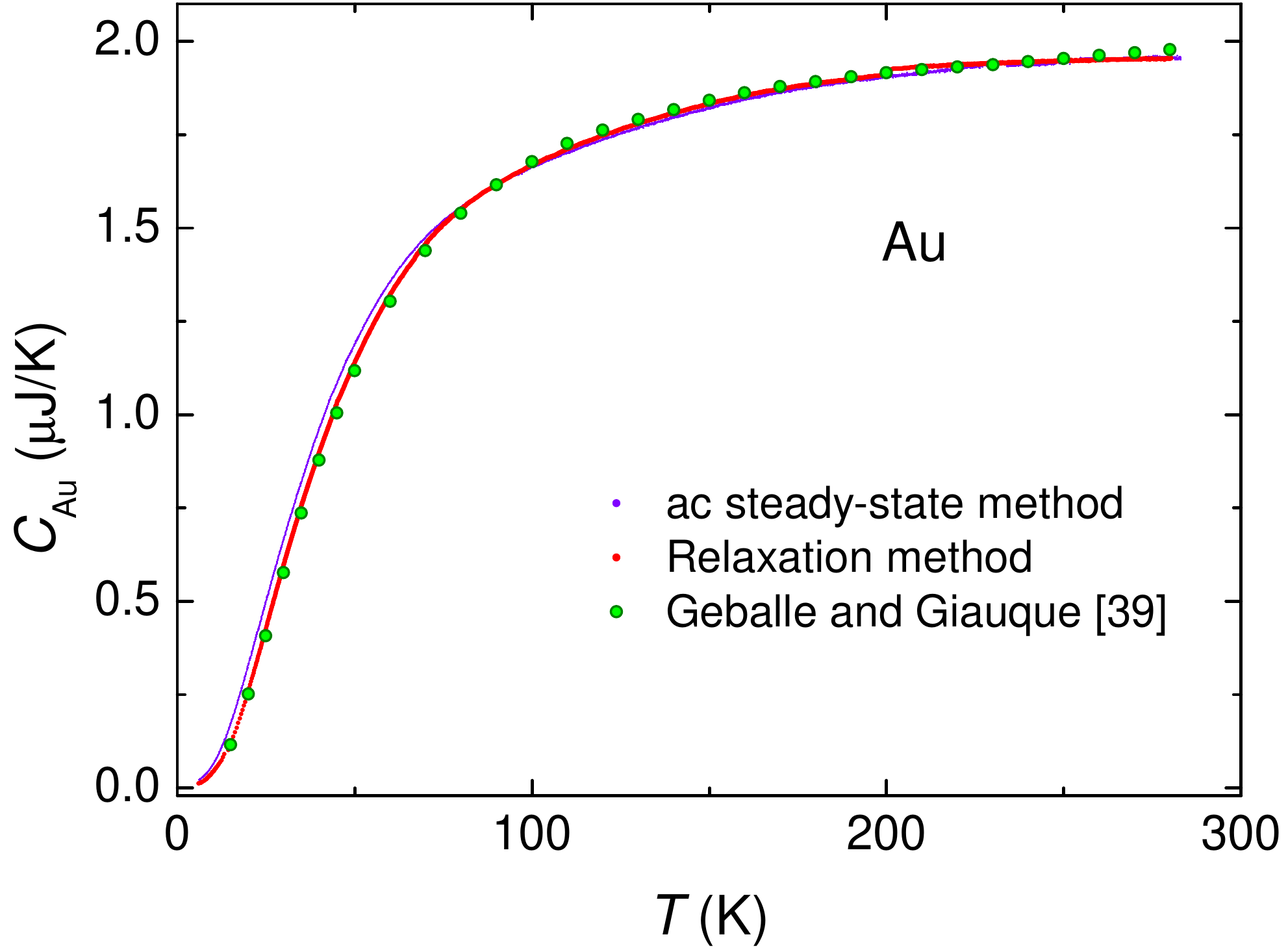}
\caption{\label{Fig8} Measured heat capacity versus temperature of a small gold grain. The contribution of the membrane heat capacity has been subtracted. There is also some contribution from the Apiezon-N grease used to attach the sample. This addenda has not been accounted for, but is expected to be of the order of $5$\,--\,$10\%$. The heat capacity measured by Geballe and Giauque\cite{Geballe:1952ct} on a sample more than $10^8$ times larger is scaled to agree with our data at $90\,\mathrm{K}$.}
\end{figure}

As an initial test of the calorimeter, we measured a small gold sample from $280\,\mathrm{K}$ down to about $10\,\mathrm{K}$ to compare the relaxation and ac steady-state techniques. The temperature dependence of the heat capacity, shown in Fig.~\ref{Fig8}, was also compared with available data to study the absolute accuracy.
It is seen that the over-all agreement between the two measurement methods is fairly good. The ac steady-state curve, however, lies somewhat above the relaxation curve for temperatures below $50\,\mathrm{K}$. The heat capacity measurement by Geballe and Giauque\cite{Geballe:1952ct} on roughly $2.5\,\mathrm{kg}$(!) Au was scaled to agree with our data at $90\,\mathrm{K}$, where the relaxation and ac steady-state measurements coincide. The relaxation curve follows the temperature dependence of the literature with deviations within 5\% over the full temperature range. From the scaling, the sample mass is estimated to $15.1\,\upmu\mathrm{g}$, which lies within the uncertainty of the volumetric measurement of the sample size, initially estimated to be $12.7\,\upmu\mathrm{g}$ using a simple microscope. The real sample mass is likely in between the two numbers, since the data of Fig.~\ref{Fig8} includes a small contribution from the Apiezon-N grease that was used to attach the sample. 

\subsection{Measurements of the superconducting properties of Pb}
\begin{figure}[t]
\includegraphics[clip,width=0.95\linewidth]{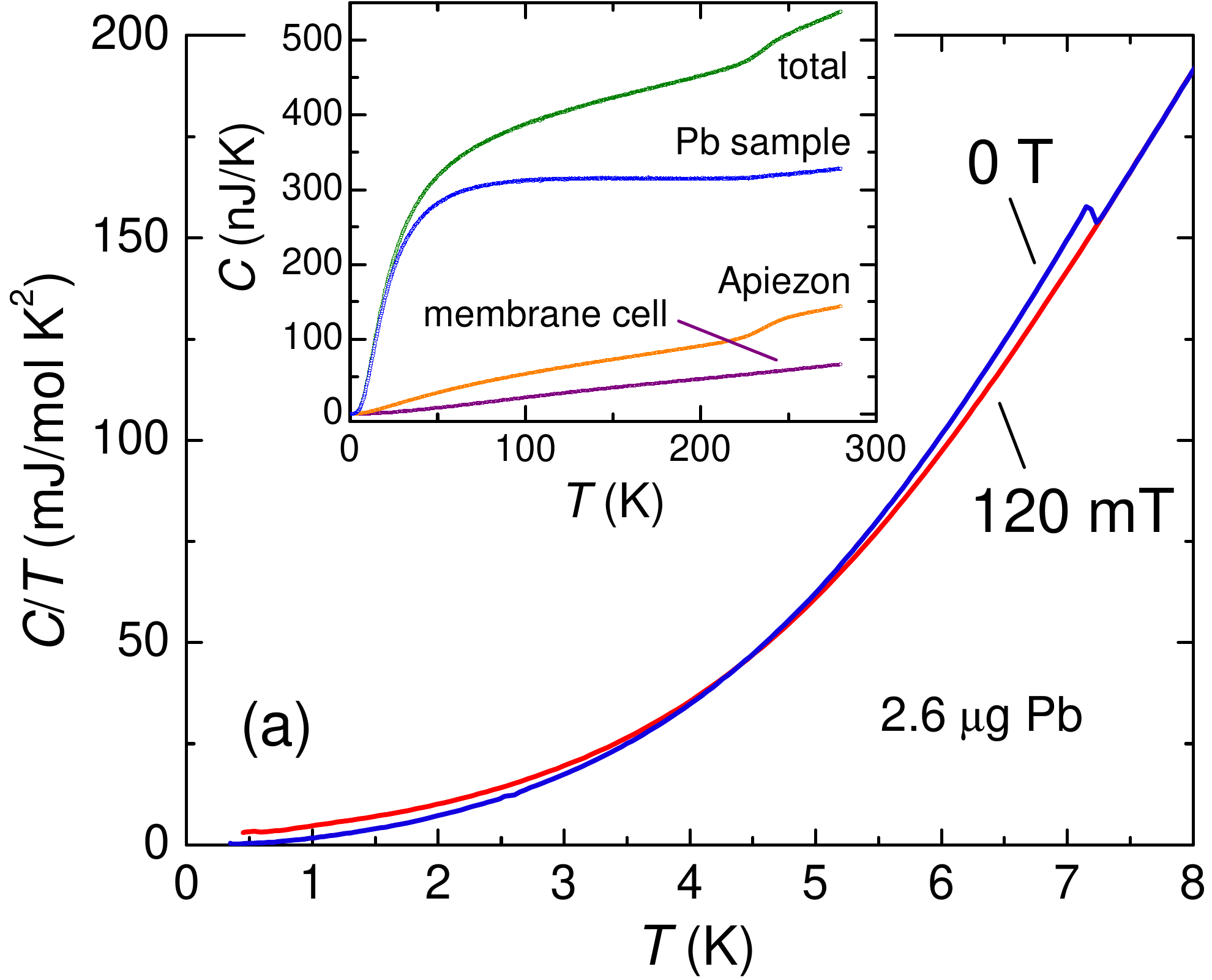}\\[1mm]
\includegraphics[clip,width=0.95\linewidth]{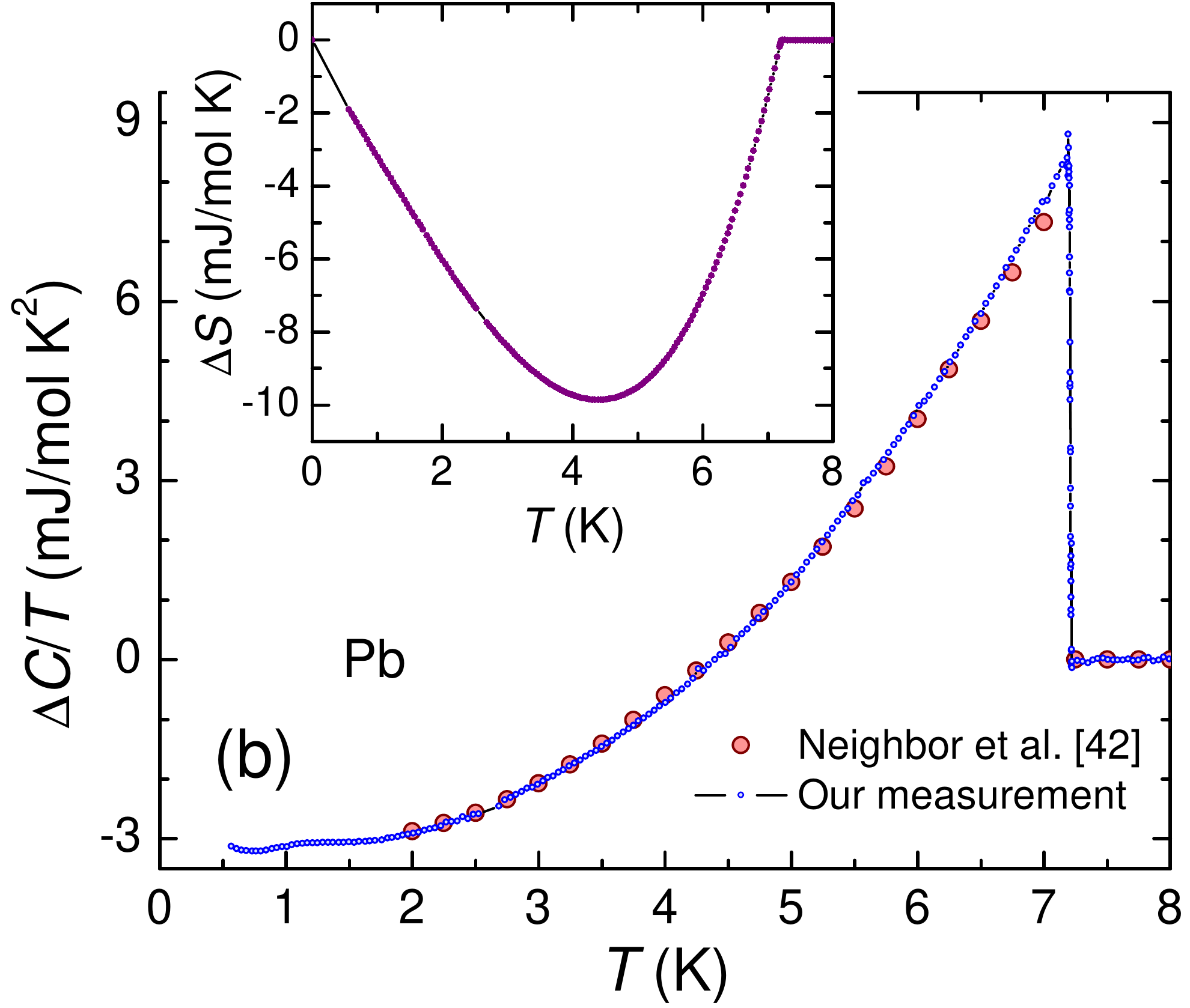}
\caption{\label{Fig9} (a) Temperature dependence of the low-temperature specific heat expressed as $C/T$ of Pb in the superconducting and normal state (obtained by a $120\,\mathrm{mT}$ field). The inset shows the measured heat capacity at higher temperatures before and after subtraction of addenda, and the contributions from membrane and Apiezon-N grease. (b) Specific heat difference $\Delta C/T = (C_\mathrm{s} - C_\mathrm{n})/T$, with the data provided by Neighbor \emph{et\,al.}\cite{Neighbor:1967fh} for comparison. The inset shows the entropy difference $\Delta S = S_s - S_n$ obtained by integrating $\Delta C/T$.}
\end{figure}

To illustrate the low-temperature capabilities of the calorimeter, we studied the heat capacity of a $2.6\,\upmu\mathrm{g}$, 99.999$\%$ pure Pb sample, cut from a single crystal, as a function of temperature and in magnetic fields. Figure~\ref{Fig9}(a) shows the measured specific heat  in the superconducting and normal states, plotted as $C/T$ vs $T$.
The specific heat jump in zero field at the  transition temperature $T_\mathrm{c} \approx 7.2\,\mathrm{K}$ is seen clearly. The normal state curve was obtained by applying a $120\,\mathrm{mT}$ magnetic field to suppress the superconductivity. The inset shows the heat capacity at higher temperatures. The addenda heat capacity from membrane and grease were measured in separate runs (i.e.\ not in true differential mode).  Apiezon-N is typically used as a thermal contact agent for low-temperature experiments, but undergoes a glass transition at above $200\,\mathrm{K}$, which leads to a somewhat irreproducible high-temperature addenda,\cite{Schnelle:1999uh,Kreitman:1972wt} decreasing the absolute accuracy at high temperatures. The membrane cell dominates the addenda at the absolute lowest temperatures, but already at about $\sim 3\,\mathrm{K}$ the Apiezon contribution becomes the main background. The membrane and grease addenda are subtracted in the main Figure~\ref{Fig9}. After subtraction, the superconducting state measurements still display a $5\%$ residual gamma term, i.e., a remaining linear-in-$T$ contribution to the specific heat. This could possibly be due to an incompletely accounted background addenda. The ratio $\Delta C/C$ at $T_\mathrm{c}$ is, however, only $95\%$ of expected. We therefore believe that the unaccounted addenda is a non-superconducting part of the sample entering $C$ but not $\Delta C$, possibly arising from an oxidized surface layer, or from vacancies and dislocations that were not annealed away before the measurements.

Figure~\ref{Fig9}(b) shows the temperature dependence of the specific heat difference $\Delta C/T = (C_\mathrm{s} - C_\mathrm{n})/T$. This difference is insensitive to background addenda, which makes it a good probe of accuracy and reproducibility. From the temperature dependence, fundamental properties such as the superconducting gap energy and coupling strength can be obtained.\cite{Carbotte:1990zz} The measurements provide a Sommerfeld parameter $\gamma \simeq 3.1\,\mathrm{mJ/mol\,K}^2$ and $\Delta C/\gamma T_\mathrm{c} \simeq 2.7$ in good agreement with literature.\cite{Carbotte:1990zz} The temperature dependence of $\Delta C/T$  is also following the expected behavior, as seen by comparing the measurements with polynomial-fit data provided by Neighbor \emph{et\,al.}\cite{Neighbor:1967fh}  While the resolution remains good at all temperatures, the accuracy decreases somewhat at $T < 1.5\,\mathrm{K}$, which can be seen in Fig.~\ref{Fig9}(b) as some wiggles in $\Delta C/T$. This can be attributed to the difficulties in obtaining an accurate calibration and corresponding sensitivity of the thermometer using $\sim\mathrm{pW}$ power levels. One way to overcome this problem may be to calibrate the thermometer simultaneously with the measurements, with a self-consistency requirement on the thermal link. Such a method has been successfully tested, but requires further investigation.

\begin{figure}[t]
\includegraphics[clip,width=0.98\linewidth]{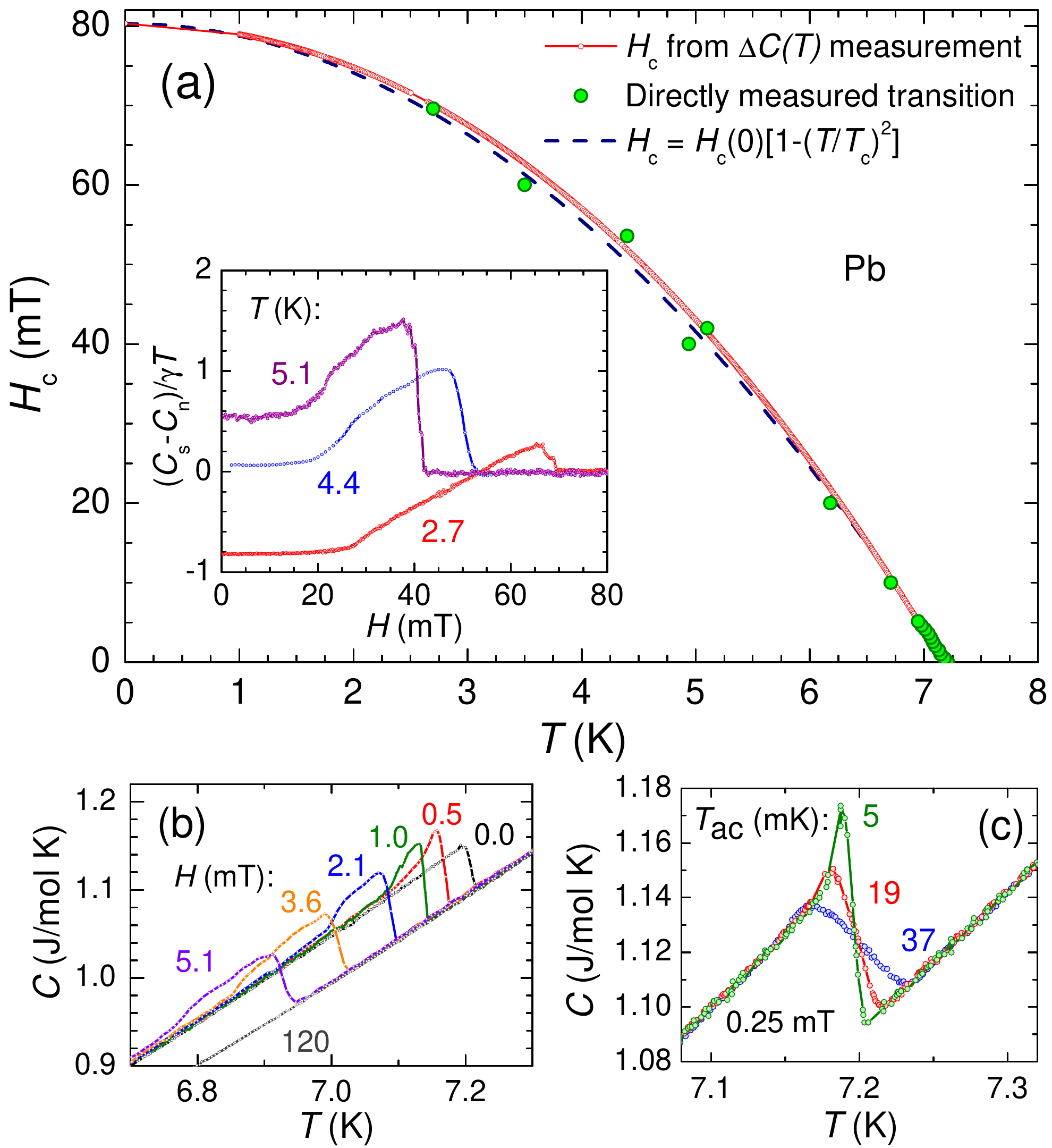}
\caption{\label{Fig10} (a) Thermodynamic critical field $H_\mathrm{c}(T)$. The small, closely spaced symbols correspond to $H_\mathrm{c}(T)$ as obtained from the $\Delta C(T)$ data of Fig.~\ref{Fig9}(b). They display the expected\cite{Neighbor:1967fh}  small deviation from the two-fluid expression, which is shown for reference. Also shown is the directly measured location of the superconducting transition in various magnetic fields (big circles). The inset shows normalized heat capacity measured on increasing fields at different temperatures. (b) Temperature dependence of the specific heat in small magnetic fields near $T_\mathrm{c}$. (c) Transition in a $0.25\,\mathrm{mT}$ magnetic field measured with different $T_\mathrm{ac}$. While improving the energy resolution, too high $T_\mathrm{ac}$ results in a pronounced $T$ smearing.}
\end{figure}

Integrating $\Delta C/T$ gives the entropy difference $\Delta S(T)$, shown in the inset of Fig.~\ref{Fig9}(b). The entropy-conservation requirement is fulfilled within a $\sim 2\%$ uncertainty of $\gamma$, obtained from the low-temperature slope of $\Delta S$. The free energy difference $\Delta F$ is then obtained as $\Delta F = \Delta U-T\Delta S$, where $\Delta U$ is found by integrating $\Delta C$ from $T_\mathrm{c}$ to $T$. From $\Delta F$, the thermodynamic critical field $H_\mathrm{c}(T)$ is calculated from the relation $\Delta F = V_\mathrm{m} \mu_0 H_\mathrm{c}^2(T)/2$, where $V_\mathrm{m}$ is the molar volume (or sample volume, if $\Delta F$ is given in units of energy). Figure~\ref{Fig10}(a) shows $H_\mathrm{c}(T)$ obtained in this way using the data in Fig.~\ref{Fig9}(b).

The temperature dependence of $H_\mathrm{c}$ for Pb is expected to display a small, positive deviation from the two-fluid expression $H_\mathrm{c}(T) = H_\mathrm{c}(0)\left[1-(T/T_\mathrm{c})^2\right]$. A weak-coupling BCS superconductor, on the other hand, would display a negative deviation.\cite{Carbotte:1990zz} The difference between $H_\mathrm{c}(T)$, as obtained from the measured $\Delta C(T)$, and the two-fluid expression is clearly seen in Fig.~\ref{Fig10}(a). Indeed, the temperature dependence of the deviations themselves are within a few \% of the deviations obtained by Decker \emph{et\,al.}\cite{Decker:1958zz,TagliatiLT26} 

The most uncertain factor in going from heat capacity to specific heat in nanocalorimetry is the determination of sample mass. It can be done through a careful measurement of volume and density, or from a known reference point at some temperature, such as room temperature or $T_\mathrm{c}$. Since a microscopic volume measurement would require sub-$\upmu\mathrm{m}$ resolution, which is difficult for soft materials such as Pb, we used the measurement of $C$ around $T_\mathrm{c}$ by Shiffman \emph{et\,al.}\cite{Shiffman:1963wz} to obtain the scale in Fig.~\ref{Fig9}. For type-I superconductors, it is however also possible to directly measure $H_\mathrm{c}(T)$ by studying the superconducting transition in magnetic field. Such measurements at various temperatures and magnetic fields are shown in the inset of Fig.~\ref{Fig10}(a) and in Fig.~\ref{Fig10}(b). The resulting $H_\mathrm{c}(T)$ and $T_\mathrm{c}(H)$ transitions, shown as big, green circles in the main panel of Fig.~\ref{Fig10}(a),  agree well with $H_\mathrm{c}(T)$ as obtained from the measurement of $\Delta C$. The sample volume can thus be found directly from a comparison of $\Delta F$ (measured in units of energy) and $\mu_0 H_\mathrm{c}^2(T)/2$ (having units of energy per volume).

The measurements in magnetic field were made with $H$ perpendicular to the plate-like sample. (The sample is shown in Fig.~\ref{Fig2}). This causes $C$ to become field dependent in the superconducting state, due to the large demagnetization factor that drives the sample into the intermediate state. The effect is clearly seen in the field-dependence curves of the inset of Fig.~\ref{Fig10}(a). At low fields the sample is in the Meissner state and $C_\mathrm{s}$ is constant, but at higher fields $C_\mathrm{s}$ starts to increase. One could interpret this effect as a distributed latent heat when normal domains enter the superconductor. In Fig.~\ref{Fig10}(b) it is seen that the specific heat in small magnetic fields is higher than the zero field specific heat near $H_\mathrm{c}$. While the sample temperature oscillates, a small fraction of the sample is undergoing the transition back and forth between the Meissner and normal states, with accompanying latent heat $T\Delta S$. This causes the $T_\mathrm{ac}$ amplitude to decrease, making the latent heat appear as an excess specific heat $C(H)-C(0) > 0$. The excess specific heat thus relates to the fraction of the sample that undergoes the superconducting transition at each temperature. It is tempting to quantify the latent heat from such measurements. However, only a fraction of the total latent heat is found in this way. This is due to  possible hysteretic effects of the first-order transition in combination with an incomplete analysis of the temperature oscillation, which will display higher harmonics when latent heat is involved.

As a final illustration of the capability of the calorimeter, the transition in a $0.25\,\mathrm{mT}$ magnetic field is shown in Fig.~\ref{Fig10}(c) for different temperature oscillation amplitudes $T_\mathrm{ac}$.  The sharp latent heat peak, which is not present in zero field  [cf.\ Fig.~\ref{Fig10}(b)], is a good probe of combined high resolution of both temperature and specific heat. The peak is only seen if $T_\mathrm{ac}$ is small enough. By increasing $T_\mathrm{ac}$ the specific heat resolution will increase, but the transition is then quickly smeared out. The latent heat involved in this transition is of the order of a few $\mathrm{pJ}$.

\section{SUMMARY AND CONCLUSIONS}
In summary, we have developed a membrane-based nanocalorimeter for specific heat measurements of small samples and thin films over an extended temperature range from above room temperature down to below $1\,\mathrm{K}$. Our device has sub-pJ/K resolution at low temperature, corresponding to $\sim 25\,\mathrm{aJ}/(\mathrm{K}\cdot\upmu\mathrm{m}^2)$ for thin films and $\sim\mathrm{fJ}$ heat exchanges, and is capable of probing $\upmu\mathrm{g}$-sized samples with combined high resolution and good absolute accuracy, thus exceeding the typical capability of commercial calorimeters by almost four orders in sample size. The calorimeter features a differential design with a variable-frequency technique where the measurement conditions are automatically maintained at optimal conditions.

The versatility of the calorimeter invites the exploration of several novel ac measurement procedures in addition to the ones described here, including power-compensation and multi-frequency modes. The ultimate capability of the calorimeter is thus still an open question.

\section{ACKNOWLEDGMENTS}
We thank S. Latos and P. A. Favuzzi for assistance with calorimetry development and R. Nilsson for contributing to the FPGA-based lock-in amplifier. Initial development of the nanocalorimeter was performed at Argonne National Laboratory in collaboration with U. Welp, W.-K. Kwok, and G. W. Crabtree. Financial support from the Swedish Research Council and technical support from the SU-Core Facility in Nanotechnology is gratefully acknowledged.

\end{document}